\documentstyle[multicol,eqsecnum,aps,psfig]{revtex}
\renewcommand{\narrowtext}{\begin{multicols}{2}
\global\columnwidth20.5pc\noindent}
\renewcommand{\widetext}{\end{multicols}
\global\columnwidth42.5pc}
\begin{document}
\draft
\preprint{22 February 2000}
\title{Modified spin-wave description of the nuclear spin relaxation
       in ferrimagnetic Heisenberg chains}
\author{Shoji Yamamoto}
\address{Department of Physics, Okayama University,
         Tsushima, Okayama 700-8530, Japan}
\date{24 February 2000}
\maketitle
\begin{abstract}
We make a modified spin-wave description of the nuclear spin
relaxation in Heisenberg alternating-spin chains with
antiferromagnetic exchange coupling.
In contrast with the conventional one-dimensional antiferromagnetic
spin-wave theory, which is plagued with the divergence of the
sublattice magnetization even in the ground state, the present
spin-wave description is highly successful over a wide temperature
range.
The temperature dependence of the relaxation rate $T_1^{-1}$
significantly varies with the crystalline structure, exhibiting both
ferromagnetic and antiferromagnetic aspects.
$T_1^{-1}$ further shows a unique dependence on the applied field,
which turns out an indirect observation of the quadratic dispersion
relations.
\end{abstract}
\narrowtext

\section{Introduction}\label{S:I}

   Quantum mixed-spin chains with magnetic ground states, namely,
quantum ferrimagnets, are one of the hot topics and recent progress
\cite
{Alca67,Tian53,Pati94,Pati07,Breh21,Ono76,Nigg31,Nigg17,Ivan24,Yama10,Yama11,Yama08,Yama33,Mais08,Kura62,Kura13,Kole65,Yama24,Wu57,Saka53,Yama75,Yama42,YamaPLA,YamaEPJB,YamaPRL}
in the theoretical understanding of them deserves special mention.
Coexistent ferromagnetic and antiferromagnetic long-range orders
\cite{Tian53} in the ferrimagnetic ground state in particular
interest us.
The ground-state magnetizations of antiferromagnets and ferromagnets
are zero and saturated, respectively, and therefore ferrimagnets may
be recognized to possess in-between ground states.
Hence the ground-state excitations in ferrimagnets are twofold
\cite{Yama10,Yama11}.
The elementary excitations of ferromagnetic aspect, reducing the
ground-state magnetization, form a gapless dispersion relation, while
those of antiferromagnetic aspect, enhancing the ground-state
magnetization, are gapped from the ground state.
The dual structure of the excitations results in unique thermal
behaviors \cite{Pati94,Pati07,Yama08,Yama33,Yama24,Wu57,YamaEPJB}:
The specific heat and the magnetic susceptibility times temperature
behave like $T^{1/2}$ and $T^{-1}$ at low temperatures, respectively,
whereas they exhibit a Schottky-like peak and a round minimum at
intermediate temperatures, respectively.
Quantum ferrimagnets in a magnetic field provide further interesting
issues such as the double-peak structure of the specific heat
\cite{Mais08,Kole65} and quantized plateaux in the ground-state
magnetization curves \cite{Kura62,Kura13,Saka53,Yama75}.
In particular it has quite recently been reported \cite{YamaPRL} that
ferrimagnetic Heisenberg chains can exhibit multi-plateau
magnetization curves even at the most symmetric point, that is,
without any anisotropy and any bond polymerization.

   It is true that theoretical investigations into quantum
ferrimagnets are now active and interesting in themselves, but we
should still be reminded that such vigorous arguments more or less
originate in the pioneering efforts \cite{Kahn89,Kahn95} to
synthesize bimetallic materials including one-dimensional systems.
The first ferrimagnetic chain compound \cite{Glei27},
MnCu(dto)$_2$(H$_2$O)$_3$$\cdot$$4.5$H$_2$O
(dto $=$ dithiooxalato $=$ S$_2$C$_2$O$_2$),
was synthesized by Gleizes and Verdaguer and stimulated the public
interest in this potential subject.
The following examples \cite{Pei38,Kahn82} of an ordered bimetallic
chain,
MnCu(pba)(H$_2$O)$_3$$\cdot$$2$H$_2$O
(pba $=$ $1,3$-propylenebis(oxamato)
 $=$ C$_7$H$_6$N$_2$O$_6$) and
MnCu(pbaOH)(H$_2$O)$_3$
(pbaOH $=$ $2$-hydroxy-$1,3$-propylenebis(oxamato)
 $=$ C$_7$H$_6$N$_2$O$_7$),
exhibiting more pronounced one dimensionality, activated further
physical \cite{Dril13,Dril92,Verd44}, as well as chemical
\cite{Pei47}, investigations.
The serial chemical explorations condensed into the crystal
engineering of a molecule-based ferromagnet \cite{Koni25}$-$the
assembly of the highly magnetic molecular entities within the crystal
lattice in a ferromagnetic fashion.

   Thus, a good amount of chemical knowledge on quasi-one-dimensional
quantum ferrimagnets has been accumulated and static properties of
them have been revealed well.
However, little is known about dynamic properties of quantum
ferrimagnets.
To the best of our knowledge, in the theoretical field, it was not
until quite recently that the dynamic structure factors were
calculated \cite{Yama11}, while in the experimental field, any direct
observation of the energy structure is not yet so successful, for
instance, as that \cite{Ma71} for the Haldane antiferromagnets
\cite{Hald64,Hald53}.
In such circumstances, Fujiwara and Hagiwara performed \cite{Fuji33}
nuclear-magnetic-resonance (NMR) measurements on bimetallic chain
compounds.
The measured temperature and applied-field ranges were rather limited
and their argument was not so conclusive.
However, they suggested that the nuclear spin relaxation could be a
useful tool in order to look into the low-energy structure peculiar
to quantum ferrimagnets.
In response to this stimulative experiment, here we calculate the
nuclear spin relaxation rate $T_1^{-1}$ in terms of a modified
spin-wave theory and strongly encourage further experimental
investigations.

\section{Formulation}\label{S:F}

   We describe alternating-spin chain compounds by the Hamiltonian
\begin{equation}
   {\cal H}
      =J\sum_{j=1}^N
        \left(
         \mbox{\boldmath$S$}_{j} \cdot \mbox{\boldmath$s$}_{j}
        +\mbox{\boldmath$s$}_{j} \cdot \mbox{\boldmath$S$}_{j+1}
        \right)
      -g\mu_{\rm B} H\sum_{j=1}^N(S_j^z+s_j^z)\,,
   \label{E:H}
\end{equation}
where
$\mbox{\boldmath$S$}_{j}^2=S(S+1)$,
$\mbox{\boldmath$s$}_{j}^2=s(s+1)$, and
we have set their $g$ factors both equal to $g$ because the
difference between them amounts to at most several per cent of
themselves in practice \cite{Hagi09}.
We further set the unit-cell length, which is twice the lattice
constant, equal to unity in the following for the convenience of
calculation.
Magnetic properties of the ferrimagnetic family compounds such as
MCu(pba)(H$_2$O)$_3$$\cdot$$2$H$_2$O and
MCu(pbaOH)(H$_2$O)$_3$ (M $=$ Mn, Ni)
are well described within this isotropic Hamiltonian
\cite{Pei38,Kahn82,Hagi09,Hagi14}.
Considering the electronic-nuclear energy-conservation requirement,
the direct (single-magnon) process is of little significance but the
Raman (two-magnon) process plays a leading role in the nuclear
spin-lattice relaxation \cite{Beem59}.
Then the relaxation rate is generally given by
\begin{eqnarray}
   &&
   \frac{1}{T_1}
    =\frac{4\pi(g\mu_{\rm B}\hbar\gamma_{\rm N})^2}
          {\hbar\sum_n{\rm e}^{-E_n/k_{\rm B}T}}
     \sum_{n,m}{\rm e}^{-E_n/k_{\rm B}T}
   \nonumber \\
   &\times&
     \big|
      \langle m|{\scriptstyle\sum_j}(A_j^zS_j^z+a_j^zs_j^z)|n\rangle
     \big|^2
     \,\delta(E_m-E_n-\hbar\omega_{\rm N})\,,
\label{E:T1def}
\end{eqnarray}
where
$A_j^z$ and $a_j^z$ are the dipolar coupling constants between
the nuclear and electronic spins in the $j$th unit cell,
$\omega_{\rm N}\equiv\gamma_{\rm N}H$ is the Larmor frequency of the
nuclei with $\gamma_{\rm N}$ being the gyromagnetic ratio, and the
summation $\sum_n$ is taken over all the electronic eigenstates
$|n\rangle$ with energy $E_n$.

   In order to rewrite the Hamiltonian (\ref{E:H}) within the
framework of the spin wave theory, we introduce the bosonic operators
for the spin deviation in each sublattice via
\begin{equation}
   \left.
   \begin{array}{lll}
      S_j^+=\sqrt{2S-a_j^\dagger a_j}\ a_j\,,&
      S_j^z=S-a_j^\dagger a_j\,,\\
      s_j^+=b_j^\dagger\sqrt{2s-b_j^\dagger b_j}\ ,&
      s_j^z=-s+b_j^\dagger b_j\,,
   \end{array}
   \right.
   \label{E:HPT}
\end{equation}
where we regard $S$ and $s$ as quantities of the same order.
Now we obtain the bosonic Hamiltonian as
\begin{equation}
   {\cal H}_{\rm SW}
    =E_{\rm class}+{\cal H}_0+{\cal H}_1+O(S^{-1})\,,
   \label{E:Hboson}
\end{equation}
where $E_{\rm class}=-2sSJN$ is the classical ground-state energy,
and ${\cal H}_0$ and ${\cal H}_1$ are the one-body and two-body terms
of the order $O(S^1)$ and $O(S^0)$, respectively.
We may consider the simultaneous diagonalization of ${\cal H}_0$ and
${\cal H}_1$ in the naivest attempt to go beyond the linear spin-wave
theory.
However, such an idea ends in failure bringing a gap to the
lowest-lying ferromagnetic excitation branch.
Thus we take an alternative approach \cite{Yama33} at the idea of
first diagonalizing ${\cal H}_0$ and next extracting relevant
corrections from ${\cal H}_1$.
${\cal H}_0$ is diagonalized as \cite{Pati94,Breh21}
\begin{equation}
   {\cal H}_0
     =E_0
     +\sum_k
      \left(
       \omega_{k}^-\alpha_k^\dagger\alpha_k
      +\omega_{k}^+\beta_k^\dagger \beta_k
      \right)\,,
   \label{E:diagH0}
\end{equation}
where
$E_0=J\sum_k[\omega_k-(S+s)]$
is the $O(S^1)$ quantum correction to the ground-state energy, and
$\alpha_k^\dagger$ and $\beta_k^\dagger$ are the creation operators
of the ferromagnetic and antiferromagnetic spin waves of momentum $k$
whose dispersion relations are given by
\begin{equation}
   \omega_{k}^\pm=\omega_k\pm(S-s)J\mp g\mu_{\rm B}H\,,
   \label{E:omegapm}
\end{equation}
with
\begin{equation}
   \omega_k=J\sqrt{(S-s)^2+4Ss\sin^2(k/2)}\,.
\end{equation}
\begin{figure}
\begin{flushleft}
\qquad\mbox{\psfig{figure=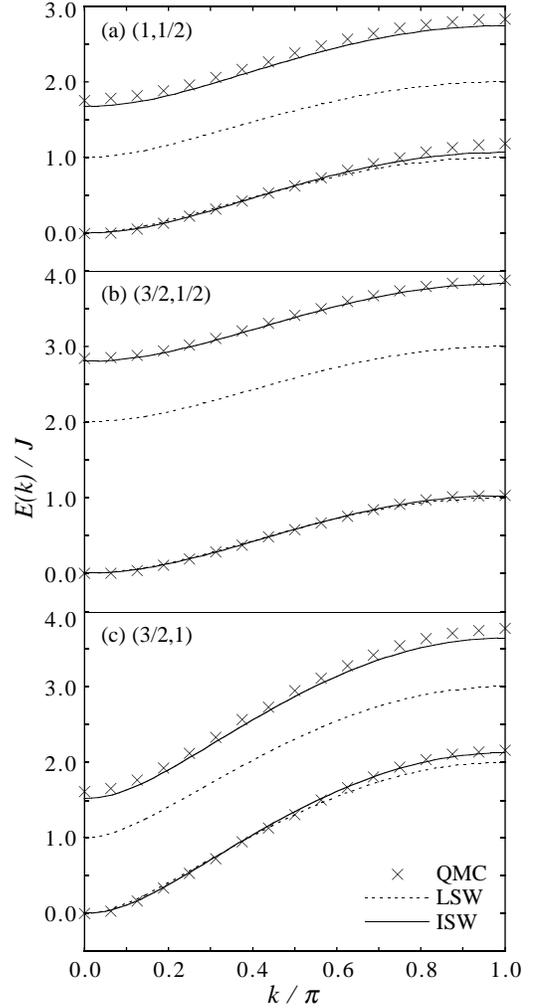,width=70mm,angle=0}}
\end{flushleft}
\vskip 0mm
\caption{Dispersion relations of the ferromagnetic and
         antiferromagnetic elementary excitations, namely, the
         lowest-energy states in the subspaces of $M=N/2\mp 1$.
         The linear- and interacting-spin-wave calculations are shown
         by the dotted and solid lines, respectively, whereas
         $\times$ represents the quantum Monte Carlo estimates
         ($N=32$).
         Here we plot the excitation energy $E(k)$ taking the
         ground-state energy as zero.}
\label{F:Ek}
\end{figure}
\noindent
The Wick theorem allows us to rewrite ${\cal H}_1$ as
\begin{eqnarray}
   {\cal H}_1
   &=&E_1-\sum_k
    \left(
    \delta\omega_k^-\alpha_k^\dagger\alpha_k
    +\delta\omega_k^+\beta_k^\dagger\beta_k
    \right)
   \nonumber\\
   &+&
     {\cal H}_{\rm irrel}+{\cal H}_{\rm quart}\,,
\end{eqnarray}
where $H_{\rm irrel}$ contains irrelevant terms such as
$\alpha_k\beta_k$ and ${\cal H}_{\rm quart}$ contains residual
two-body interactions, both of which are neglected in the following
so as to keep the low-energy structure qualitatively unchanged.
$E_1=-2JN
    [{\mit\Gamma}_1^2+{\mit\Gamma}_2^2
    +(\sqrt{S/s}+\sqrt{s/S})
     {\mit\Gamma}_1{\mit\Gamma}_2]$
is the $O(S^0)$ correction to the ground-state energy, while
\begin{equation}
   \delta\omega_k^\pm
      =2(S+s){\mit\Gamma}_1
       \frac{\sin^2(k/2)}{\omega_k}
      +\frac{{\mit\Gamma}_2}{\sqrt{Ss}}
       [\omega_k\pm(S-s)]\,,
    \label{E:domega}
\end{equation}
are those to the dispersions, where the key constants
${\mit\Gamma}_1$ and ${\mit\Gamma}_2$ are defined as
${\mit\Gamma}_1
  =(2N)^{-1}\sum_k
   [(S+s)/\omega_k-1]$ and
${\mit\Gamma}_2
  =-N^{-1}\sum_k
   (\sqrt{Ss}/\omega_k)\cos^2(k/2)$.
The resultant Hamiltonian is compactly represented as
\begin{equation}
   {\cal H}_{\rm SW}
     \simeq E_{\rm g}
    +\sum_k
     \left(
      {\widetilde\omega}_k^- \alpha_k^\dagger \alpha_k
     +{\widetilde\omega}_k^+ \beta_k^\dagger  \beta_k
     \right)\,,
   \label{E:HSW}
\end{equation}
with $E_{\rm g}=E_{\rm class}+E_0+E_1$ and
${\widetilde\omega}_k^\pm=\omega_k^\pm-\delta\omega_k^\pm$.

   We show in Fig. \ref{F:Ek} the linear- and interacting-spin-wave
dispersions, $\omega_k^\pm$ and ${\widetilde\omega}_k^\pm$, together
with the numerical solutions \cite{YamaEPJB} obtained through
imaginary-time quantum Monte Carlo calculations
\cite{Yama48,Yama45,Yama09}.
The spin-wave description of the low-energy structure is fairly good.
Even the linear spin waves allow us to have a qualitative view of the
elementary excitations.
The relatively poor description of the antiferromagnetic branch by
the linear spin waves reminds us of the spin-wave treatment of
mono-spin Heisenberg chains, where the theory accurately describes
ferromagnetic chains, while it only gives a qualitative view of
antiferromagnetic chains.
The spin-wave approach to the present system is highly successful
anyway for both excitation branches.
The spin-wave series potentially lead to the goal even for the
antiferromagnetic branch.
The high applicability essentially originates in the fact that the
spin deviations
\begin{eqnarray}
   &&
    \frac{1}{N}\sum_j\langle a_j^\dagger a_j\rangle_{\rm g}
   =\frac{1}{N}\sum_j\langle b_j^\dagger b_j\rangle_{\rm g}
   ={\mit\Gamma}_1
   \nonumber \\
   &&\quad
   =\frac{1}{2\pi}\int_0^\pi
    \biggl[
     \frac{S+s}{\sqrt{(S-s)^2+4Ss\sin^2(k/2)}}-1
    \biggr]{\rm d}k\,,
\end{eqnarray}
with $\langle\ \ \rangle_{\rm g}$ denoting the ground-state average,
no more diverge in the present system with $S\neq s$.
We are convinced that the quantity ${\mit\Gamma}_1$ should be
recognized as the quantum spin reduction.

   In terms of the spin waves, the relaxation rate (\ref{E:T1def}) is
expressed as
\begin{eqnarray}
   \frac{1}{T_1}
   &=&\frac{4\pi\hbar}{N^2}(g\mu_{\rm B}\gamma_{\rm N})^2
      \sum_{k,q}\sum_{\sigma=\pm}
      \delta(\omega_{k+q}^\sigma-\omega_k^\sigma-\hbar\omega_{\rm N})
   \nonumber \\
   &\times&
   \bigl[
    (A_q^z{\rm cosh}\theta_{k+q}{\rm cosh}\theta_k)^2
    n_k^\sigma(n_{k+q}^\sigma+1)
   \nonumber \\
   &&\!\! +
    (a_q^z{\rm sinh}\theta_{k+q}{\rm sinh}\theta_k)^2
    n_{k+q}^\sigma(n_k^\sigma+1)
   \nonumber \\
   &&\!\! -
    2A_q^z a_q^z({\rm cosh}\theta_k{\rm sinh}\theta_k)^2
    n_k^\sigma(n_k^\sigma+1)
   \bigr]\,,
   \label{E:T1SW}
\end{eqnarray}
where
$n_k^-\equiv\langle\alpha_k^\dagger\alpha_k\rangle$ and
$n_k^+\equiv\langle \beta_k^\dagger\beta_k \rangle$ are the thermal
averages of the boson numbers at a given temperature, and
$A_q^z=\sum_j{\rm e}^{{\rm i}q(j-1/4)}A_j^z$ and
$a_q^z=\sum_j{\rm e}^{{\rm i}q(j+1/4)}a_j^z$
are the Fourier components of the hyperfine coupling constants.
Taking into account the significant difference between the electronic
and nuclear energy scales ($\hbar\omega_{\rm N}\alt 10^{-5}J$), Eq.
(\ref{E:T1SW}) ends in
\begin{eqnarray}
   \frac{1}{T_1}
   &=&\frac{4\hbar}{NJ}(g\mu_{\rm B}\gamma_{\rm N})^2
      \sum_k\frac{S-s}{\sqrt{(Ssk)^2+2(S-s)Ss\hbar\omega_{\rm N}/J}}
   \nonumber \\
   &\times&
   \bigl[
    (A^z{\rm cosh}^2\theta_k-a^z{\rm sinh}^2\theta_k)^2
    n_k^-(n_k^- +1)
   \nonumber \\
   &&\!\! +
    (A^z{\rm sinh}^2\theta_k-a^z{\rm cosh}^2\theta_k)^2
    n_k^+(n_k^+ +1)
   \bigr]\,,
   \label{E:T1final}
\end{eqnarray}
where we have assumed little $k$-dependence of $A_q^z$ and $a_q^z$,
and thus replaced $A_{q=-2k}^z$ and $a_{q=-2k}^z$ by
$A_{q=0}^z\equiv A^z$ and $a_{q=0}^z\equiv a^z$, respectively.

\begin{figure}
\begin{flushleft}
\ \ \mbox{\psfig{figure=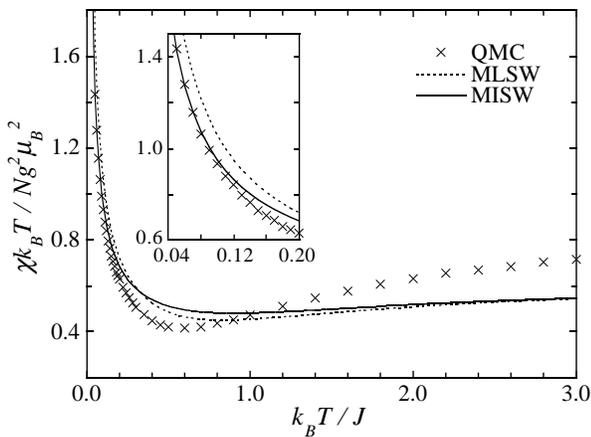,width=80mm,angle=0}}
\end{flushleft}
\vskip 0mm
\caption{Temperature dependences of the zero-field magnetic
         susceptibility times temperature in the case of
         $(S,s)=(1,\frac{1}{2})$.
         $\times$ represents the quantum Monte Carlo estimates,
         whereas the dotted and solid lines show the
         modified-spin-wave calculations starting from the
         linear- and interacting-spin-wave dispersion relations,
         respectively.
         The low-temperature behaviors are scaled up in the inset.}
\label{F:chiT}
\end{figure}

   The estimation of the relaxation rate is now reduced to the
calculation of the spin-wave distribution functions.
Though the ground-state distribution is well controlled, the naivest
thermodynamics, based on the partition function
$Z={\rm Tr}[{\rm e}^{-{\cal H}_{\rm SW}/k_{\rm B}T}]$,
breaks down as temperature increases.
Hence we modify the spin-wave theory
\cite{Taka68,Taka94,Hirs69,Tang00} introducing an additional
constraint in minimizing the free energy.
Requiring the total magnetization to be zero, Takahashi \cite{Taka68}
obtained an excellent description of the low-temperature
thermodynamics of one-dimensional Heisenberg ferromagnets.
Taking his core idea but replacing the ferromagnetic constraint
$\sum_j\langle S_j^z+s_j^z\rangle=0$ by
\begin{equation}
   \sum_j
    \bigl(
     \langle S_j^z-s_j^z\rangle+2{\mit\Gamma}_1
    \bigr)
    =(S+s)
     \biggl(
      N-\sum_k\sum_{\sigma=\pm}\frac{n_k^\sigma}{\omega_k}
     \biggr)
    =0\,,
   \label{E:const}
\end{equation}
we obtain the modified spin-wave distribution functions as
\begin{equation}
   n_k^\pm
    =\frac{1}
     {{\rm e}^{[{\widetilde\omega}_k^\pm-\mu(S+s)J/\omega_k]
               /k_{\rm B}T}-1}\,,
\end{equation}
with a Lagrange multiplier $\mu$ due to the condition
(\ref{E:const}).
In comparison with the ferromagnetic cases, the quantum correction
$2{\mit\Gamma}_1$ is necessary for ferrimagnets.
The spin-wave treatment gives
$\langle S_j^z+s_j^z\rangle_{\rm g}=S+s$
under ferromagnetic exchange coupling but
$\langle S_j^z-s_j^z\rangle_{\rm g}=S+s-2{\mit\Gamma}_1$
under antiferromagnetic exchange coupling.
Indeed, without the quantum correction, we reach a quite poor
description of the thermal quantities \cite{Yama08}.
We show in Fig. \ref{F:chiT} the modified spin-wave calculations of
the magnetic-susceptibility-temperature product, which is closely
related with the relaxation rate \cite{Hone65} and is given in terms
of $n_k^\pm$ as
\begin{equation}
   \chi T
    =\frac{(g\mu_{\rm B})^2}{3k_{\rm B}}
     \sum_k\sum_{\sigma=\pm}
     n^\sigma_k(n^\sigma_k+1)\,.
   \label{E:chiT}
\end{equation}
The obtained results are fairly successful considering that these are
the spin-wave calculations in one dimension.
The modified spin waves well reproduce the low-temperature
ferromagnetic divergence, which is proportional to $T^{-1}$, the
high-temperature antiferromagnetic increase toward the paramagnetic
behavior $[S(S+1)+s(s+1)]/3$, and therefore, the round minimum at
intermediate temperatures.
In particular, the low-temperature description by the interacting
spin waves may be regarded as accurate.
Now we proceed to the argument of the relaxation rate in terms of the
modified interacting-spin-wave theory.
Since the applied field $H$ is so small in practice \cite{Fuji33}
as to satisfy $g\mu_{\rm B}H\alt 10^{-2}J$, we neglect the Zeeman
term of the Hamiltonian (\ref{E:HSW}) in the estimation of $n_k^\pm$.

\section{Results}\label{S:R}

   Here still remains an adjustable parameter $A^z/a^z\equiv r$.
The dipolar coupling is quite sensitive to the location of the
nuclei because the coupling strength is proportional to $d^{-3}$ with
$d$ being the distance between the interacting nuclear and electronic
spins.
In the proton-NMR measurements on NiCu(pba)(H$_2$O)$_3$$\cdot$2H$_2$O
\cite{Fuji33}, for instance, it was demonstrated that the protons
relevant to the relaxation rate do not originate in the H$_2$O
molecules but
\begin{figure}
\begin{flushleft}
\ \ \mbox{\psfig{figure=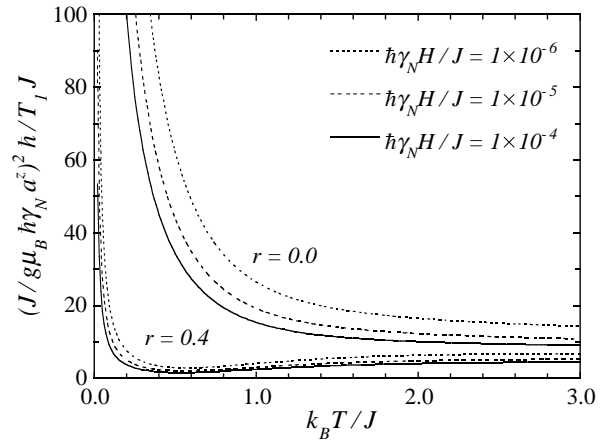,width=80mm,angle=0}}
\end{flushleft}
\vskip -3mm
\caption{Temperature dependences of the nuclear spin relaxation rate
         at various values of the applied magnetic field in the case
         of $(S,s)=(1,\frac{1}{2})$.}
\label{F:T1T112}
\end{figure}
\vskip -1mm
\begin{figure}
\begin{flushleft}
\qquad\mbox{\psfig{figure=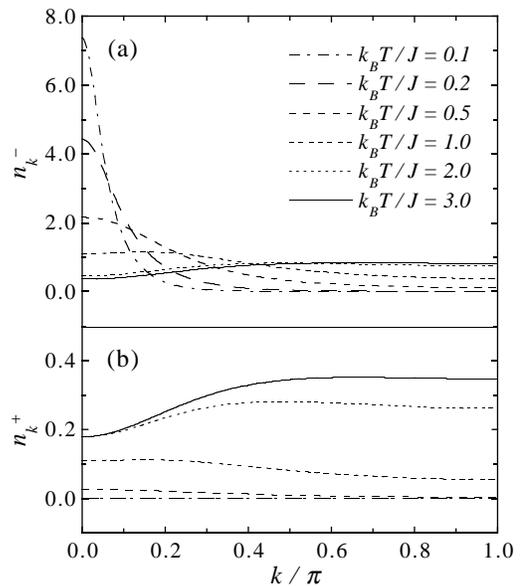,width=70mm,angle=0}}
\end{flushleft}
\vskip -3mm
\caption{The momentum distribution functions of the ferromagnetic (a)
         and antiferromagnetic (b) spin waves in the modified
         spin-wave theory starting from the interacting-spin-wave
         dispersion relations in the case of
         $(S,s)=(1,\frac{1}{2})$.}
\label{F:nkMISW}
\end{figure}
\vskip 4mm
\noindent
lie in the pba groups which are located beside the Cu
ions.
Thus, for these family compounds, $r$ may reasonably be set equal to
zero.
We show in Fig. \ref{F:T1T112} the corresponding calculations.
As the measurements were performed at rather high temperatures, which
are beyond the quantitative reliability of the present theory, it is
impossible to fit the calculations to the experimental findings.
However, the calculations at $r=0$ well explain the observations of
monotonically decreasing behaviors of $T_1^{-1}$ as a function of
temperature, which are in contrast with the appearance of $\chi T$.
When we compare Eqs. (\ref{E:T1final}) and (\ref{E:chiT}), we realize
that $T_1^{-1}$ could tell more than $\chi T$ due to its adjustable
prefactors to $n_k^\sigma(n_k^\sigma+1)$.
Setting $r$ equal to $0.4$, we obtain temperature dependences of
$T_1^{-1}$ exhibiting a round minimum, where the antiferromagnetic
spin-wave contribution $n_k^+$ is much more accentuated than the
ferromagnetic one $n_k^-$.
Let us observe the momentum dependences of $n_k^\pm$ in Fig.
\ref{F:nkMISW}.
$n_k^-$ exhibits a sharp peak around $k=0$ at low temperatures,
which is rapidly reduced with the increase of temperature.
On the other hand, $n_k^+$ is an increasing function of temperature
at an arbitrary momentum, though its broad peak is smeared out with
the increase of temperature.
The field-dependent prefactor in Eq. (\ref{E:T1final}), coming from
the energy conservation requirement
$\delta(E_m-E_n-\hbar\omega_{\rm N})$ in Eq. (\ref{E:T1def}),
predominantly extracts the $k=0$ components from $n_k^\pm$.
Therefore, the parametrization
$r=0.4\simeq\frac{1}{2}={\rm tanh}^2\theta_{k=0}$
results in the strong suppression of the ferromagnetic aspect of
$T_1^{-1}$.
\begin{figure}
\begin{flushleft}
\ \ \mbox{\psfig{figure=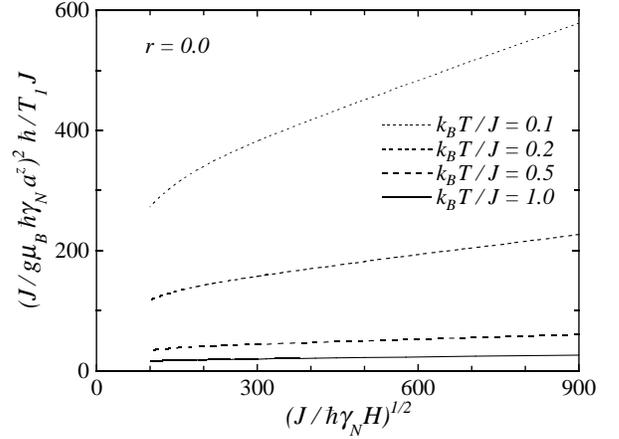,width=80mm,angle=0}}
\end{flushleft}
\vskip -2mm
\caption{Field dependences of the nuclear spin relaxation rate
         at various values of temperature in the case of
         $(S,s)=(1,\frac{1}{2})$.}
\label{F:T1H112}
\end{figure}
\vskip 0mm
\widetext
\begin{figure}
\begin{flushleft}
\qquad\quad\mbox{\psfig{figure=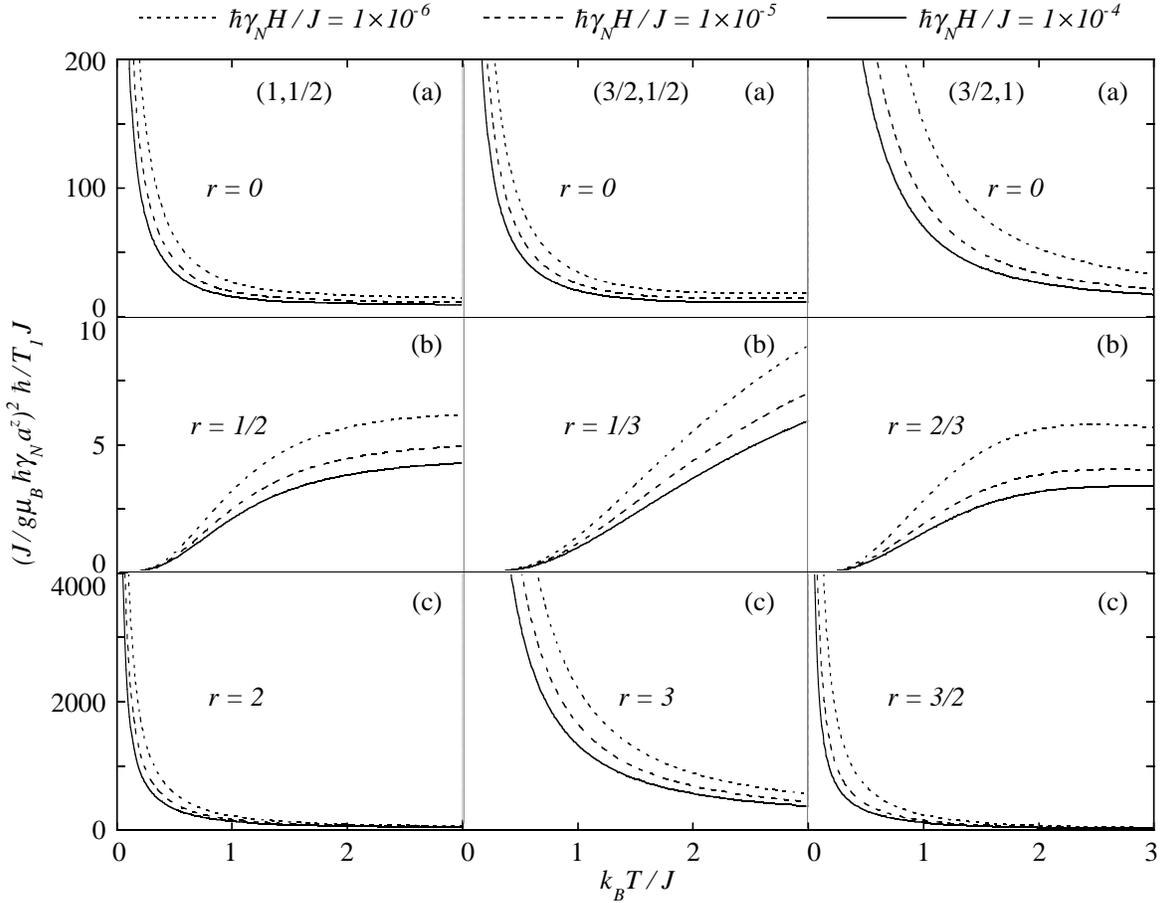,width=156mm,angle=0}}
\end{flushleft}
\vskip 0mm
\caption{Temperature dependences of the nuclear spin relaxation rate
         at various values of the applied magnetic field in various
         constituent-spin cases.}
\label{F:T1Tgen}
\end{figure}
\narrowtext

   Another motivation of the NMR measurements on ferrimagnetic chains
may be the characteristic field dependences of $T_1^{-1}$ shown in
Fig. \ref{F:T1H112}.
It is due to the quadratic dispersion relations (\ref{E:omegapm})
that $T_1^{-1}$ depends on the applied field.
Hence the present field dependence allows us to look into the
low-energy structure peculiar to ferrimagnets.
The high linearity of $T_1^{-1}$ with respect to $H^{-1/2}$ denotes
the predominance of the $k\simeq 0$ components in the $k$-summation
in Eq. (\ref{E:T1final}).
The predominance is reduced with the increase of $H$ and finally
there arises a logarithmic field dependence of $T_1^{-1}$ from the
$k$-integration of
$[(Ssk)^2+2(S-s)Ss\hbar\omega_{\rm N}/J]^{-1/2}$.
If we take $J/k_{\rm B}=121[\mbox{K}]$ \cite{Hagi09} relevant to
NiCu(pba)(H$_2$O)$_3$$\cdot$$2$H$_2$O, Fig. \ref{F:T1H112} suggests
that the logarithmic behavior should appear under $H\agt 10$[T].
The $T_1$ measurements \cite{Fuji33} on
NiCu(pba)(H$_2$O)$_3$$\cdot$$2$H$_2$O at $280$[K] with an applied
field up to $3.15$[T] resulted in a monotonic linear dependence of
$T_1^{-1}$ on $H^{-1/2}$.
The high-field measurements at lower temperatures are expected.

   In order to stimulate extensive NMR measurements on ferrimagnetic
compounds, we show in Fig. \ref{F:T1Tgen} temperature dependences of
$T_1^{-1}$ in general constituent-spin cases.
We select three particular values of $r$:
(a) $r=0$, 
where the relevant nuclei are located closely to the smaller spins
$s$ and the ferromagnetic and antiferromagnetic mixed nature is
displayed;
(b) $r={\rm tanh}^2\theta_{k=0}$,
where the relevant nuclei are located near the smaller spins $s$
rather than the larger spins $S$ and the ferromagnetic aspect is
strongly suppressed; and
(c) $r={\rm coth}^2\theta_{k=0}$, 
where the relevant nuclei are located near the larger spins $S$
rather than the smaller spins $s$ and the antiferromagnetic aspect
is strongly suppressed.
Thus, the chemical technique might enable us to extract separately
the ferromagnetic/antiferromagnetic feature of ferrimagnets from
$T_1^{-1}$.
Let us observe the temperature dependences from the point of view of
the low-energy structure, which has been revealed in Fig. \ref{F:Ek}.
The distinct behaviors Fig. \ref{F:T1Tgen}(b) and Fig.
\ref{F:T1Tgen}(c) are reminiscent of the antiferromagnetic and
ferromagnetic $\chi T$ products, respectively.
The antiferromagnetic aspect should be dominated by the
antiferromagnetic gap $\widetilde\omega_{k=0}^+$.
Although the spin-wave description within the up-to-$O(S^1)$
approximation considerably underestimates the antiferromagnetic gap,
its estimate $\omega_{k=0}^+=2(S-s)J$ can be a qualitative guide.
We learn that the antiferromagnetic gap is in proportion to $S-s$ and
indeed find the slower activation and the higher-located peak of the
antiferromagnetic component of $T_1^{-1}$ for
$(S,s)=(\frac{3}{2},\frac{1}{2})$ in comparison with those for
$(S,s)=(1,\frac{1}{2})$ and $(S,s)=(\frac{3}{2},1)$.
A careful observation of Fig. \ref{F:T1Tgen}(a) shows that $T_1^{-1}$
reaches a minimum around $k_{\rm B}T/J=2.5$ for
$(S,s)=(\frac{3}{2},\frac{1}{2})$.
In the two other cases, $T_1^{-1}$ is still on the decrease at
$k_{\rm B}T/J=6.0$.
We may understand whether $T_1^{-1}$ exhibits a minimum$-$that is to
say, a ferromagnetic-to-antiferromagnetic crossover$-$at a tractable
temperature in connection with the balance between the ferromagnetic
band width $W^-$ and the antiferromagnetic gap ${\mit\Delta}$.
The ferromagnetic decreasing tail exists for $k_{\rm B}T\alt W^-$,
whereas the antiferromagnetic increasing behavior is remarkable for
$k_{\rm B}T\alt{\mit\Delta}$.
If we evaluate $W^-$ and ${\mit\Delta}$ as
$\omega_{k=\pi}^- - \omega_{k=0}^- = 2Js$ and
$\omega_{k=0}^+ = 2(S-s)J$, respectively, we may expect a detectable
minimum of $T_1^{-1}$ as a function of temperature for $2s<S$.
The larger $S-2s$, the more pronounced crossover may be detected.

\section{Summary and Discussion}\label{S:SD}

   A relaxation mechanism based on the interaction with spin waves
has so far been applied to magnetic insulators only in a temperature
range far below the onset of the long-range order \cite{Beem59}.
We again stress that the present argument is a positive use of the
spin-wave theory in one dimension in a temperature range above the
onset of the (three-dimensional) long-range order.
The quantum divergence of the spin reduction, inherent in
one-dimensional antiferromagnets, no more plagues the present system,
while we have avoided the thermal divergence of the bosonic
distribution function modifying the spin-wave theory, that is,
imposing a certain constraint on the magnetization.
Though the present analysis must be a qualitative guide for the
experiments over a wide temperature range, the description may be
very precise especially for $k_{\rm B}T/J\alt 0.2$.
$T_1$ measurements at low temperatures and/or under high fields are
strongly encouraged.

   The experimental development depends on the synthesis of relevant
materials.
Although it is the pioneering measurements \cite{Fuji33} on
NiCu(pba)(H$_2$O)$_3$$\cdot$2H$_2$O that have motivated the present
study, the family compounds MCu(pba)(H$_2$O)$_3$$\cdot$2H$_2$O
may not be so useful as to verify the present analysis.
The linewidth in the NMR spectra for
NiCu(pba)(H$_2$O)$_3$$\cdot$2H$_2$O considerably broadens at low
temperatures and therefore the extraction of $T_1$ was restricted to
a temperature range $k_{\rm B}T/J\agt 0.5$, where it is rather hard
for the present theory to make a quantitative interpretation, namely,
a precise determination of the geometric parameters $A^z$ and $a^z$.
Even though the linewidth broadening is inevitably dominated by the
crystalline structure, a strong exchange coupling must be desirable
here anyway.
The idea of designing ligands capable of binding to two different
metal ions with different donor atoms has not yet ended in an
exchange coupling constant $J$ beyond
$85$[cm$^{-1}$] $\simeq$ $122k_{\rm B}$[K] \cite{Kahn95}.
A breakthrough may be made by bringing into interaction metal ions
and stable organic radicals.
Ceneschi {\it et al.} \cite{Cane91,Cane56,Cane76,Cane40,Cane14}
synthesized a series of ferrimagnetic chain compounds of general
formula M(hfac)$_2$NITR, where metal ion complexes M(hfac)$_2$ with
hfac $=$ hexafluoroacetylacetonate are bridged by nitronyl nitroxide
radicals NITR.
Their exchange coupling constants significantly vary with M and R but
are in general fairly large in comparison with those of bimetallic
chain compounds.
An antiferromagnetic coupling of
$313$[cm$^{-1}$] $\simeq$ $450k_{\rm B}$[K] was obtained for
M $=$ Mn ($S=\frac{5}{2}$) and R $=$ isopropyl ($s=\frac{1}{2}$)
\cite{Cane76}, while that of
$424$[cm$^{-1}$] $\simeq$ $610k_{\rm B}$[K] for
M $=$ Ni ($S=1$) and R $=$ methyl ($s=\frac{1}{2}$) \cite{Cane40}.
Neither compound shows any transition to the three-dimensional
magnetic order down to $k_{\rm B}T/J=0.02$, which is also suitable
for our purpose, apart from designing a molecule-based ferromagnet.
We theoreticians hope as many experimentalists as possible will take
part in this exciting business$-$dynamic properties of
one-dimensional quantum ferrimagnets.

\acknowledgments

   It is a pleasure to thank Prof. T. Fukui and Dr. N. Fujiwara for
fruitful discussions.
This work was supported by the Japanese Ministry of Education,
Science, and Culture through Grant-in-Aid No. 11740206 and by the
Sanyo-Broadcasting Foundation for Science and Culture.
The numerical computation was done in part using the facility of the
Supercomputer Center, Institute for Solid State Physics, University
of Tokyo.

\widetext
\end{document}